\documentclass[aps,prb,superscriptaddress,twocolumn]{revtex4-1}
\usepackage{graphicx}% Include figure files
\usepackage{dcolumn}% Align table columns on decimal point
\usepackage{bm}% bold math
\usepackage{amsmath}
\usepackage[colorlinks = true,
            linkcolor = red,
            urlcolor  = violet,
            citecolor = blue,
            anchorcolor = violet]{hyperref}
\usepackage[table]{xcolor}
\usepackage{array} 

\definecolor{darkgreen}{rgb}{0.0,0.4,0.0}

\newcommand{\tc}{T$_{\textmd c }$}
\newcommand{\Tc}{T$_{\textmd c }$}

\newcommand{\omlog}{$\omega_{\textmd l}$}
\newcommand{\chin}{\chi_{\text{\vspace{-0.12cm}\hspace{-0.03cm}\scriptsize 0}}}
\newcommand{\MPIHalle}{Max-Planck Institut f\"ur Microstrukturphysik, Weinberg 2, 06120 Halle, Germany} 
\newcommand{\rdc}[1]{\scriptsize{#1}}

\begin{document}

\title{Calculations of critical temperatures in conventional superconductors: beyond the random phase approximation}
\title{Coulomb interactions and conventional superconductivity: when going beyond the random phase approximation is essential}

\author{Camilla Pellegrini} \email{pellegrini.physics@gmail.com}
\affiliation{\MPIHalle}

\author{Carl Kukkonen} \email{Kukkonen@Cox.net}
\affiliation{33841 Mercator Isle, Dana Point, California 92629, USA}
%\affiliation{We envy you}

\author{Antonio Sanna}
\email{sanna@mpi-halle.mpg.de}
\affiliation{\MPIHalle}

\date{\today}

\begin{abstract}
In ab initio calculations of superconducting properties, the Coulomb repulsion is accounted for at the GW level and is usually computed in RPA, which amounts to neglecting vertex corrections both at the polarizability level and in the self-energy. Although this approach is unjustified, the brute force inclusion of higher order corrections to the self-energy is computationally prohibitive. We propose to use a generalized GW self-energy, where vertex corrections are incorporated into W by employing the Kukkonen and Overhauser (KO) ansatz for the effective interaction between two electrons in the electron gas. By computing the KO interaction in the adiabatic local density approximation for a diverse set of conventional superconductors, and using it in the Eliashberg equations, we find that vertex corrections lead to a systematic decrease of the critical temperature (\Tc),
ranging from a few percent in bulk lead to more than 40\% in compressed lithium.  
We propose a set of simple rules to identify those systems where large \Tc\ corrections are to be expected and hence the use of the KO interaction is recommended. Our approach offers a rigorous extension of the RPA and GW methods for the prediction of superconducting properties at a negligible extra computational cost.

\end{abstract}

\pacs{Valid PACS appear here}
\maketitle

\section{Introduction}

The established approach to calculate the transition temperature (\Tc) of conventional superconductors within Eliashberg theory~\cite{Eliashberg_JETP,AllenMitrovic1983,ScalapinoSchriefferWilkins_StrongCouplingSC_PR1966} relies on a GW-like approximation for both the phonon-mediated and the Coulomb part of the electron self-energy. 
For the electron-phonon case the validity of this approximation is supported by Migdal's theorem, that states that vertex corrections are negligible if the phonon energy scale, set by the Debye frequency $\omega_D$, is much smaller than the electronic Fermi energy $E_F$.
However, there is no small parameter that enables a simplified perturbative treatment of the Coulomb interaction between the electrons. In this latter case, the GW approach~\cite{AryasetiawanGunnarson_GW_RepProgPhys1998,VonsovskySuperconductivityTransitionMetals} is an unjustified approximation, and an accurate description of Coulomb effects in the superconducting state would require including vertex corrections both at the polarizability level and in the self-energy.

By setting the vertex function equal to one, the GW approximation is given by the self-consistent electron Green's function times the screened Coulomb potential
\begin{equation}
W^{tt}(q,\omega)=\frac{v_q}{\varepsilon(q,\omega)}=v_q+v^2_q \chi_{nn}(q, \omega),
\label{eq:tt}
\end{equation}
where $v_q=4\pi e^2/q^2$ is the bare Coulomb potential, $\varepsilon(q,\omega)$ is the dielectric function and $\chi_{nn}(q,\omega)$ is the density-density response function. 
Importantly, Eq.~(\ref{eq:tt}) describes the interaction between two external test charges embedded in the many-body medium: it includes the bare exchange interaction, as well as the screening of all the interactions stemming from the rearrangement of the electronic charge in response to the addition of a test particle to the system. In practice, $W$ is commonly evaluated in the random phase approximation (RPA), which also neglects exchange and correlation (xc) contributions to the electron polarizability~\cite{FetterWalecka_QuantumTheoryOfManyPatricleSystems_Book1971,GiulianiVignale_ElectronLiquid_Book2005}. 

Apparently, any attempt to improve over the standard GW scheme is hampered by the huge numerical complexity of computing higher order corrections to the electron self-energy.
To overcome this problem, in this work we propose to use a generalized GW self-energy, where vertex corrections are absorbed into the definition of an effective interaction W. One such interaction was obtained phenomenologically by Kukkonen and Overhauser (KO)~\cite{Kukkonen_elelInteractionSimpMetals_PRB1979}, and later derived by Vignale and Singwi using diagrammatic techniques~\cite{GiulianiVignale_ElectronLiquid_Book2005}. Unlike Eq.~(\ref{eq:tt}), the KO interaction is a realistic model for the interaction between two physical electrons in the homogeneous electron gas (HEG). It includes xc effects within the medium, but also recognizes that the two electrons that are to be paired for superconductivity are identical to the electrons in the screening cloud, so that xc effects between that Cooper pair and the rest of the system are also included. All many-body effects or, equivalently, vertex corrections to the polarization and the self-energy, are conveniently incorporated into the KO interaction in a local approximation by making use of local field factors that define the density and spin response functions of the HEG.

The purpose of the present paper is to examine the influence of vertex corrections on the \Tc\ of real materials by employing the KO ansatz for the effective electron-electron interaction. For the computation of \Tc\ we resort to a recent implementation of the Eliashberg equations, that allows for a computationally efficient ab initio treatment of the Coulomb interactions\cite{Pellegrini_SimpEliashberg_JOPM2022}.

\section{The Kukkonen Overhauser effective $\text{el-el}$ interaction}\label{sec:II} 
The Eliashberg equation for the pairing function $\phi(k,i\omega_n)$ can be written as:
\begin{align}
 \phi(k,i\omega_n)=&\frac{1}{\beta}\sum_{k',n'}\Bigl[\frac{\lambda(k,k';i\omega_n-i\omega_{n'})}{N(0)}\nonumber\\
 &- \tilde{I}_{s}(k,k';i\omega_n,i\omega_{n'})\Bigr]\frac{\phi(k',i\omega_{n'})}{\Theta(k',i\omega_{n'})},
 \end{align}
 where $\Theta=\left(i\omega_{n}Z\right)^2-\varepsilon^2_{\boldsymbol{k}}-\phi^2$.
Here, the first term accounts for the phonon exchange, where $\lambda(k,k';i\omega_n-i\omega_{n'})$ is the electron-phonon coupling and $N(0)$ is the electron density of states at the Fermi level. $\tilde{I}_{s}(k,k';i\omega_n,i\omega_{n'})$ is defined as the irreducible electron-electron interaction for the scattering of a pair in the singlet state with momenta and Matsubara frequencies $(k,i\omega_n;-k,-i\omega_n)$ to the final state with $(k',i\omega_{n'};-k',-i\omega_{n'})$.
 In order to utilize a proper form of $\tilde{I}_{s}(k,k';i\omega_n,i\omega_{n'})$ that includes xc effects, we resort to the model proposed by KO for the effective interaction $W_{\sigma \sigma'}(q,\omega)$, which describes the scattering of two electrons with spins $\sigma$ and $\sigma'$ for momentum and energy transfer $(q,\omega)$ in the HEG.
In this scheme, $W^{KO}(q, \omega)$ is composed of the bare interaction $v(q)$ and the interactions mediated by charge and spin density fluctuations:
\begin{align}
 W^{KO}(q, \omega)=&v_q +\{v_q \left[ 1-G_+(q,\omega)\right] \}^2 \chi_{n n}(q,\omega)\nonumber\\
 &-3\{v_q
 G_-(q,\omega)\}^2\chi_{S_zS_z}(q,\omega)\label{eq:KO}.
\end{align}
These latter contributions are constructed from the charge and spin dynamical local-field factors $G_{\pm}(q, \omega)$, that include xc effects. Note that the coefficient -3 in front of the spin fluctuation term comes from the assumption of spin singlet pairing. $\chi_{n n}\left(q,\omega\right)$ and $\chi_{S_zS_z}\left(q,\omega\right)$ are, respectively, the charge and spin response functions defined by
\begin{align}
 &\chi_{n n}\left(q,\omega\right)=\frac{\chin\left(q,\omega\right)}{1-v_q\left[1-G_{+}\left(q,\omega\right)\right]\chin\left(q,\omega\right)}\label{eq:X00_eg}\\
  &\chi_{S_zS_z}\left(q,\omega\right)=\frac{\chin\left(q,\omega\right)}{1+v_qG_{-}\left(q,\omega\right)\chin\left(q,\omega\right)},\label{eq:Xzz_eg}
\end{align}
in terms of the free-electron response function $\chi^0(q, \omega)$ and the local field factors $G_{\pm}(q, \omega)$, as well. 
By neglecting in Eq.~(\ref{eq:KO}) exchange and correlation between the two interacting electrons and the medium, while keeping them within the medium, one recovers the test particle-test particle interaction of Eq.~(\ref{eq:tt}).
As already mentioned, $W^{tt}(q, \omega)$ is usually computed in RPA, which amounts to setting $G_+(q, \omega)=0$ in $\chi_{n n}\left(q,\omega\right)$ (Eq.(\ref{eq:X00_eg})), entirely discarding xc effects.
We anticipate that our numerical analysis will show that an accurate approximation to the full KO interaction for the prediction of \Tc\ is given by the following expression: 
\begin{equation}
W^{KO^+}\left(q,\omega\right)=v_q+v_q^2\left[1-2G_+\left(q,\omega\right)\right]\chi_{n n}\left(q,\omega\right), \label{eq:linKO}
\end{equation}
which has the computational advantage of not depending on $\chi_{S_zS_z}$, that is a quantity not always available in linear response codes.

In this work we do not concern ourselves with anisotropy effects on \Tc, and rely on the Eliashberg equations in the isotropic limit derived in Refs.~\onlinecite{Pellegrini_SimpEliashberg_JOPM2022,Adavydov_PRB2020,Sanna_Eliashberg_JPSJ2018}. Within this approach, the electron-phonon coupling is described by the Eliashberg spectral function $\alpha^2F\,$~\cite{AllenMitrovic1983}, and the screened Coulomb interaction $W_{k,k'}$  is approximated by its average over surfaces of constant energy ($\varepsilon$) in $k$-space, that is,
\begin{equation}
W(\varepsilon,\varepsilon')=\frac{1}{N(\varepsilon)N(\varepsilon')} \sum_{k,k'} W_{k,k'}\delta(\varepsilon_k-\varepsilon)\delta(\varepsilon_{k'}-\varepsilon'),\label{eq:Waverage}
\end{equation} 
where $N(\varepsilon)=\sum_k \delta(\varepsilon_k-\varepsilon)$ is the density of electronic states. 
For later convenience we also introduce the dimensionless quantity $\mu=W(0,0)N(0)$, that enters the simplified Morel-Anderson scheme~\cite{MorelAnderson_1962} for Coulomb renormalization. This is given by the product between the average Coulomb interaction at the Fermi level ($\varepsilon=0$) and the Fermi density of states. It should be clear that $\mu$ does not enter our simulations, which employ the full function $W(\varepsilon,\varepsilon')$. Nevertheless, it will serve in the discussion as a rough estimate of the Coulomb repulsion strength at the Fermi level, where it is physically more meaningful.

\subsection{Calculations for the homogeneous electron gas}\label{sec:eg}

As a first step, we compare the strength of the Coulomb interaction in the different approximation schemes for the HEG at varying density parameter (Wigner-Seitz radius) $r_s$. 
By making the system superconducting with the addition of a coupling to an Einstein phonon mode (with frequency $\omega=60$ meV and strength $\lambda=1$), we then compute the corresponding Eliashberg critical temperatures.

\begin{figure}[t] 
\begin{center}
\includegraphics[width=0.9\columnwidth]{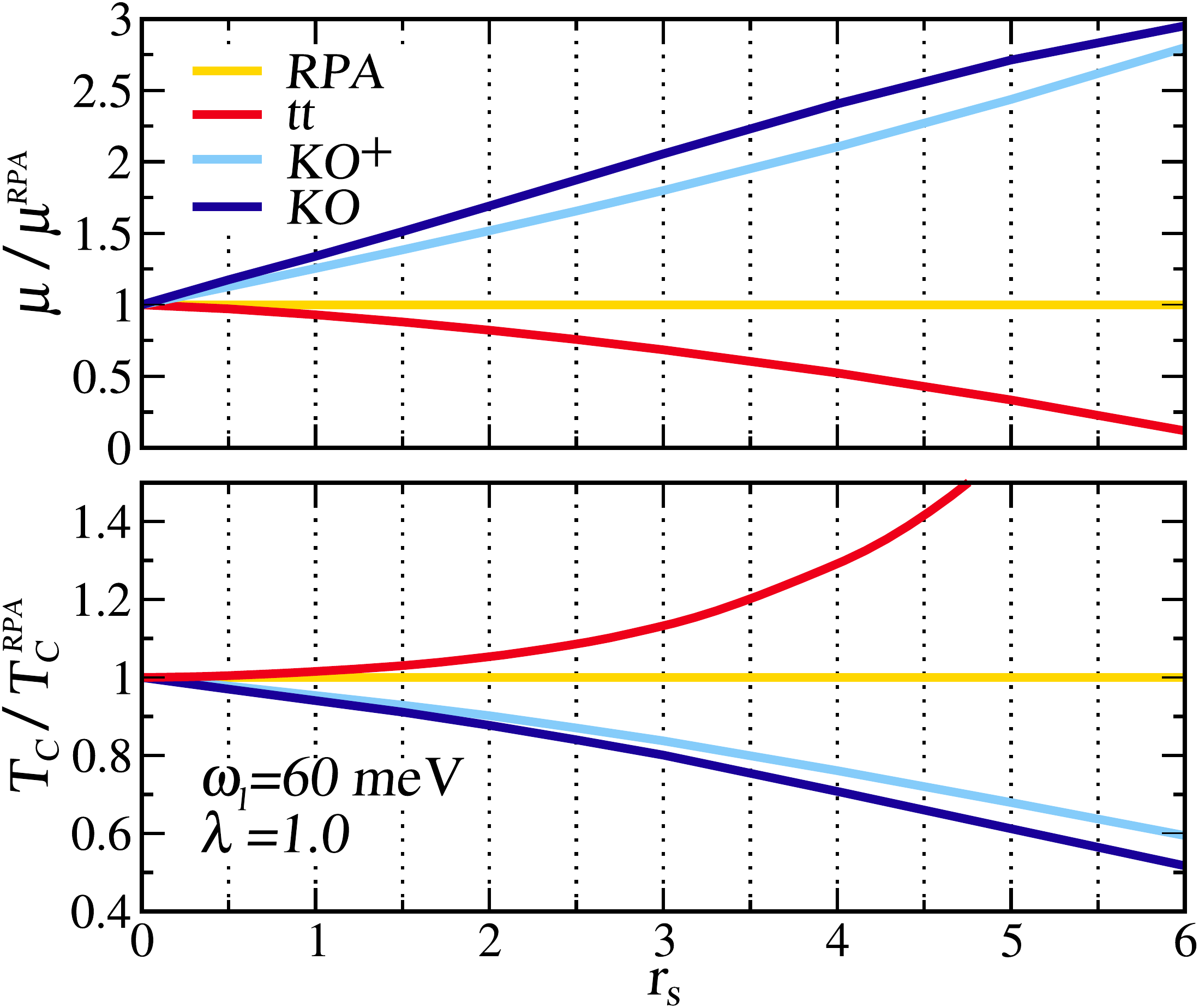} 
\caption{Impact of xc correlation effects on the Coulomb repulsion (top) and critical temperature (bottom) in the homogeneous electron gas. The RPA contains no xc effect. tt is the test particle test particle interaction. KO is the singlet Kukkonen Overhauser effective interaction in Eq.~\ref{eq:KO}. KO$^+$ is an approximated form of the KO interaction  defined in Eq.~\ref{eq:linKO}.\\
The attractive part of the superconducting pairing is provided by an Einstein phonon mode with coupling $\lambda$=1 and frequency \omlog=60meV. All results are given as a function of the gas density expressed by $r_s$.
}
\label{fig:W_electronGas}
\end{center}
\end{figure}

In the upper panel of Fig.~\ref{fig:W_electronGas} we plot the ratio of $\mu$ as computed from $W^{tt}$, $W^{KO}$ and $W^{KO^+}$ [Eqs.~(\ref{eq:tt}), (\ref{eq:KO}) and (\ref{eq:linKO})] divided by $\mu^{RPA}$.
For the local-field factors we have taken the simple quadratic expressions recently proposed by Kukkonen and Chen~\cite{KukkonenChen_EffectiveInteractionPRB2021}:
\begin{align}
G_+(q)&=\left(1-\frac{k_0}{k}  \right)\left(\frac{q}{q_{\textsubscript{TF}}}\right)^2,\\
G_-(q)&=\left(1-\frac{\chi_0}{\chi}  \right)\left(\frac{q}{q_{\textsubscript{TF}}}\right)^2,
\end{align}
where the compressibility and susceptibility ratios are $r_s$-dependent.
These expressions are exact at small $q$, and accurately reproduce quantum Monte Carlo data up to $q=2k_F$ within the metallic region $r_s=1-5$.

We observe that for typical metallic densities ($r_s=2-3$) the effective KO repulsion at the Fermi level is stronger than the RPA by about a factor of 2, whereas the static screening of the bare interaction is the most effective in the test particle-test particle approximation.
 
The corresponding critical temperatures as a function of the density are reported in the lower panel of Fig.~\ref{fig:W_electronGas}. 
As expected from the values of W, the RPA overestimates \Tc\ with respect to the KO approximation. The calculations for this model system indicate that the inclusion of vertex corrections lowers the \tc\ by about 20\% at conventional metallic densities.
\textcolor{black}{A recent paper by one of the authors used the KO interaction to calculate the superconducting parameters ($\mu$ and $\lambda$), and reached the same conclusion that the repulsive parameter is increased compared to when the RPA is used. Thus the superconducting transition temperature calculated using the McMillan formula is reduced\cite{Kukkonen_PRB2023}.}
However, as we will see by studying real materials (Sec.~\ref{sec:AnalysisTcMaterials}), fine details of the electronic and vibrational properties, that are neglected in the present model, can considerably affect this result.

\newcommand{\nMgB}{\footnote{Accounting for anisotropy \Tc\ would increase to 34K~\cite{Pellegrini_SimpEliashberg_JOPM2022}}}
\newcommand{\srp}{R\!P\!A}
\newcommand{\slk}{l\!K\!O}
\newcommand{\sko}{K\!O}

\newcommand\ChangeRT[1]{\noalign{\hrule height #1}}
\begin{table*}
\begin{tabular}{l!{\vrule width 0.8pt} c | c | c | c !{\vrule width 0.8pt} c | c | c | c !{\vrule width 1pt} c | c | c | c } 
    & $\mu^{RPA}$  & $\Delta\mu^{tt}$\rdc{(\%)}& $\Delta\mu^{KO^+}$\rdc{(\%)}&$\Delta\mu^{KO}$\rdc{(\%)}& \Tc$^{RPA}$\rdc{(K)} & $\Delta$\Tc$^{tt}$\rdc{(\%)}& $\Delta$\Tc$^{KO^+}$\rdc{(\%)}&$\Delta$\Tc$^{KO}$\rdc{(\%)}& $r_s$ & $\alpha^2\!F$ & $\lambda$ &\omlog\rdc{ (meV)} \\
     \ChangeRT{0.8pt}\hline 
%                    mu     DmuALDA DmulKO DmuKO    TcEli  DTcALDA DTclKO  DTcKo   rs
Al               & 0.206 & -20.1 & +39.4 & +49.5 &  1.03 & +17.0 & -32.7 & -41.9 &2.48&\onlinecite{SPG_EliashbergSCDFT_PRL2020}&0.37&24.6\\\hline
In               & 0.208 & -15.8 & +34.0 & +41.0 &  4.27 & +3.5  & -6.9  &  -8.3 &3.12&\onlinecite{SPG_EliashbergSCDFT_PRL2020}&0.84& 6.2\\\hline
C:(5\%B)         & 0.151 &  -9.2 & +13.4 & +16.9 &  3.93 & +14.1 & -17.0 & -20.5 &3.83&\onlinecite{Boeri_Diamond_PRL2004}      &0.36&122.6\\\hline
Li(@22GPa)       & 0.487 & -51.0 & +58.7 & +83.8 &  7.82 & +28.2 & -30.1 & -43.0 &1.22&\onlinecite{Profeta_LiAlK_PRB2006}      &0.77&22.9\\\hline
Pb               & 0.239 & -15.6 & +33.8 & +40.6 &  6.85 & +2.1  & -4.1  &  -4.9 &2.77&\onlinecite{SPG_EliashbergSCDFT_PRL2020}&1.33& 5.1\\\hline
RbSi$_{\text{2}}$& 0.268 & -31.0 & +44.6 & +60.2 & 10.1  & +7.1  & -8.7  & -12.4 &3.42&\onlinecite{Livas_graphites_PRB2015}    &1.28& 8.6\\\hline
Nb               & 0.515 & -17.9 & +34.9 & +47.6 & 12.4  & +4.8  & -8.1  & -11.0 &0.50&\onlinecite{SPG_EliashbergSCDFT_PRL2020}&1.34&12.0\\\hline
P(@15GPa)        & 0.178 & -11.4 & +23.0 & +29.1 & 13.0  & +2.2  & -3.5  & -4.47 &4.24&\onlinecite{Livas_Ppressure_PRM2017}    &1.04&13.4\\\hline
NbSe$_{\text{2}}$& 0.501 & -21.8 & +31.7 & +43.0 & 11.6  & +8.0  & -10.5 & -14.5 &1.04&\onlinecite{Sanna_NbSe2_npjQM2022}      &1.43&12.0\\\hline 
MgB$_{\text{2}}$ & 0.265 & -17.7 & +32.5 & +42.7 &18.6\nMgB &+11.1&-19.1 & -26.1 &2.02&\onlinecite{SPG_EliashbergSCDFT_PRL2020}&0.67&61.3\\\hline
Nb$_{\text{3}}$Sn& 0.589 & -26.1 & +40.4 & +58.1 & 20.7  & +7.9  & -11.1 & -17.2 &0.49&\ref{app:Nb3Sn}                         &1.86&14.5\\\hline
SH$_{\text{3}}$  & 0.220 & -12.0 & +27.7 & +35.8 & 211.1 & +3.0  & -6.8  &  -9.2 &1.34&\onlinecite{Errea_SH3_PRL2015}          &1.90&91.6\\
\ChangeRT{0.8pt}\hline
set average      & 0.319 & -20.8 & +34.8 & +45.7 &  26.8 & +9.1  & -13.2 & -17.8 &2.21&                                        &1.10&32.9\\
\end{tabular}  

\caption{Ab initio critical temperatures and Coulomb parameters, for a selected set of conventional superconductors. $\mu^{RPA}$ and T$_{\textmd c }^{RPA}$ are the Coulomb repulsion parameter (at the Fermi level) and the critical Temperature, computed within the RPA approximation. $\Delta\mu$ and $\Delta$\Tc\ are the percent variation of these quantities by adopting the improved effective interaction  test particle-test particle ($tt$), the Kukkonen-Overhauser (KO) interaction and its simplified form (KO$^+$). Citations indicate the source of the electron phonon interaction used for these tests, of which we report the integrated total coupling ($\lambda$) and the logarithmic averaged frequency (\omlog). }\label{tab:data}
\end{table*}

%%%%%%%%%%%%%%%%%%%%%%%%%%%%%%%%%%%%%%%%%%%%%%%%%%%%%%%%%%%%%%%%%%%%%%%%%%%%%
\section{Application to real materials}

\subsection{The KO interaction for inhomogeneous systems}\label{sec:KOinhomogeneous}

The HEG is the basis of the large majority of approximations to the xc functionals of density-functional theory in materials science.
In TD-SDFT calculations, a common approximation to the unknown xc functional of real (inhomogeneous) systems involves the adiabatic kernel,
\begin{equation}
 f^{\sigma\sigma'}_{\text{xc}}( n^{GS}_{\alpha},|\mathbf{r}-\mathbf{r}'|)  =\left.\frac{\delta^2 E_{xc}[n_{\uparrow}, n_{\downarrow}]}{\delta n_{\sigma}(\mathbf{r}) \delta n_{\sigma'}(\mathbf{r}')}\right |_{n_{\alpha}=n^{GS}_{\alpha}},
\end{equation}
where $E_{xc}[n_{\uparrow}, n_{\downarrow}]$ is the xc energy of the HEG with ground-state spin densities $n^{GS}_{\alpha}$. In the paramagnetic state, the spin-symmetric and spin-antisymmetric xc kernels $f^{\pm}_{xc}(r)$ are simply related to the local-field factors $G^{\pm}(q)$ by the Fourier transform:
\begin{align}
 f^{\pm}_{xc}(q)&\equiv \frac{f^{\uparrow\uparrow}_{xc}(q)\pm f^{\uparrow\downarrow}_{xc}(q)}{2}=\int d\mathbf{r} e^{-i\mathbf{q}\cdot \mathbf{r}} f^{\pm}_{xc}(r)  \nonumber\\
 &=-v_q G^{\pm}(q).
 \end{align}

For lattice periodic systems, the KO interaction of Eq.~(\ref{eq:KO}) reads as:
\begin{align}
W_{{\bf G}{\bf G}'}\left({\bf q}\right)=&\frac{4\pi \delta_{{\bf G}{\bf G}'}}{\left|{\bf q}+{\bf G}\right|^2} \nonumber\\
                   +&\sum_{{\bf G}_1{\bf G}_2}\bigg[ f_{Hxc,\mathbf{G}\mathbf{G}_1}\left({\bf q}\right) f_{Hxc,\mathbf{G}_1\mathbf{G}_2}\left({\bf q}\right)\chi^{nn}_{{\bf G}_2{\bf G}'}({\bf q}) \nonumber\\
                   -& 3f^-_{xc,\mathbf{G}\mathbf{G}_1}\left({\bf q}\right)f^-_{xc,\mathbf{G}_1\mathbf{G}_2}\left({\bf q}\right)\chi^{S_zS_z}_{{\bf G}_2{\bf G}'}({\bf q})\bigg],\label{eq:KO_periodic}
\end{align}
where we have defined the Hartree-xc kernel $$f_{Hxc,{\bf G}{\bf G}'}\left({\bf q}\right)=\frac{4\pi \delta_{{\bf G}{\bf G}'}}{\left|{\bf q}+{\bf G}\right|^2}+f^{+}_{xc,{\bf G}{\bf G}'}\left({\bf q}\right).$$
The interacting density-density and spin-spin response functions entering Eq.~(\ref{eq:KO_periodic}) can be obtained, in principle exactly, from the Dyson-like equations:
\begin{align}
 \chi^{nn}_{\mathbf{G} \mathbf{G}'}(\mathbf{q})&= \chi^{\text{KS}}_{\mathbf{G} \mathbf{G}'}(\mathbf{q})
 +\sum_{\mathbf{G}_1,\mathbf{G}_2}\chi^{\text{KS}}_{\mathbf{G} \mathbf{G}_1}(\mathbf{q})\nonumber\\
 &\times f_{Hxc,\mathbf{G}_1\mathbf{G}_2}(\mathbf{q}) \chi^{nn}_{\mathbf{G}_2 \mathbf{G}'}(\mathbf{q}),\label{eq:chinn}\\
 \chi^{S_zS_z}_{\mathbf{G} \mathbf{G}'}(\mathbf{q})&= \chi^{\text{KS}}_{\mathbf{G} \mathbf{G}'}(\mathbf{q})
 +\sum_{\mathbf{G}_1,\mathbf{G}_2}\chi^{\text{KS}}_{\mathbf{G} \mathbf{G}_1}(\mathbf{q})\nonumber\\
 &\times f^-_{xc,\mathbf{G}_1\mathbf{G}_2}(\mathbf{q})  \chi^{S_zS_z}_{\mathbf{G}_2 \mathbf{G}'}(\mathbf{q}),\label{eq:chispinspin}
 \end{align}
where $\chi^{\text{KS}}_{\mathbf{G} \mathbf{G}'}(\mathbf{q})$ is the Kohn-Sham response function. In practice, the xc kernels $f^{\pm}_{xc,\mathbf{G}\mathbf{G}'}$ are usually computed in the adiabatic local density approximation (ALDA), which amounts to replacing the static kernel of the HEG by its long-wavelength limit. This value is then used at each point in space according to the local density of the system, i.e.,
\begin{equation}
f^{\pm,\text{ALDA}}_{xc,\mathbf{G}\mathbf{G}'}=-\lim_{q\to 0}\frac{1}{\Omega}\int_{\text{cell}}d\mathbf{r}\,\, e^{-i(\mathbf{G}-\mathbf{G}')\cdot \mathbf{r}}\, 
v_q G^{\pm}(q, r_s(\mathbf{r})).
\end{equation}
 
We have checked that the (ALDA) xc kernels computed from the local-field factors of Ref.~\onlinecite{KukkonenChen_EffectiveInteractionPRB2021} are almost identical  
to those routinely calculated in TDDFT from the second functional derivative of the HEG xc energy, when adopting the Perdew and Wang parametrization for the correlation energy. For computational convenience, we have thus evaluated the KO interaction in real materials by using in Eqs.~(\ref{eq:chinn}) and ~(\ref{eq:chispinspin}) the ALDA Perdew-Wang xc kernels as calculated with the Elk code.

Since ab initio superconductivity calculations are carried out in the basis of the Kohn-Sham orbitals, Eq.~(\ref{eq:KO_periodic}) has been implemented in the form
\begin{equation}
W_{k,k'}=\frac{1}{\Omega}\sum_{{\bf G}{\bf G}'}W_{{\bf G}{\bf G}'}\left({\bf q}\right)\rho^k_{k'}\left({\bf G}\right)\rho^{k\,*}_{k'}\left({\bf G}'\right),
\end{equation}
where $\rho^k_{k'}({\bf G})= \left< k'|e^{-i({\bf q}+{\bf G})\cdot {\bf r}}|k\right >$, $k$ stands for the band index $n$ and momentum ${\bf k}$ of the Kohn-Sham state and ${\bf q}\equiv{\bf k}-{\bf k}'$.

%%%%%%%%% MATERIALS %%%%%%%%%%%%%%%
\subsection{The material test set}\label{sec:MaterialSet}

From the complete solution of the isotropic Eliashberg equations including both effective Coulomb and electron-phonon interactions, we have calculated the superconducting transition temperatures of a diverse set of conventional superconductors.  
The materials in the set have been chosen so as to cover a wide range of properties and conditions (under which Coulomb effects are expected to play a significant role), and hence they also include exotic superconductors. This implies that comparing with experimental results will not always be straightforward or possible. We have considered elemental superconductors such as Al, In, Pb and Nb. Al~\cite{Profeta_LiAlK_PRB2006} is a prototype weak-coupling superconductor with a very low critical temperature (1.2~K). Since its electronic structure is nearly free-electron, it is expected to behave similarly to the HEG. In passing from Al to In, Pb and Nb, the electronic charge becomes gradually more localized, the electron-phonon coupling increases, and so does \Tc~\cite{Floris_Pb_PRB2007,SPG_EliashbergSCDFT_PRL2020}. Among the elemental superconductors, we have also included lithium under pressure~\cite{Shimizu_LiPressure_Nature2002,Profeta_LiAlK_PRB2006,Akashi_LithiumPlasmons_PRL2013}. This system becomes superconducting owing to a s$\to$p charge transfer, hence its electronic behavior is at the crossing point between free electrons and more localized charge carriers. 
Additionally, we have considered two Nb compounds, Nb$_3$Sn and NbSe$_2$. Nb$_3$Sn is one of the most relevant superconductors for high-field generation applications, as it features high critical fields and a relatively high \Tc\ of 18~K~\cite{ShiraneAxe_NeutronNb3Sn_PRB1971}. NbSe$_2$ is a layered superconductor (\Tc=7.2K) made famous by the coexistence of superconductivity and charge density wave~\cite{Sanna_NbSe2_npjQM2022}. We have added to our set three more layered superconductors, RbSi$_2$, an (hypotetical) intercalated silicate with honeycomb structure~\cite{LivasSanna_honeycombs_PRB2015}, black phosphorus at high pressure~\cite{Livas_Ppressure_PRM2017} and magnesium diboride~\cite{Nagamatsu_MgB2_Nature2001,Floris_MgB2_PRL2005,Floris_MgB2_PhysicaC2007}. Layered materials usually display stronger Coulomb repulsion because of the inherent charge localization, as compared to three dimensional systems~\cite{Pellegrini_SimpEliashberg_JOPM2022}. Compressed black phosphorus~\cite{Livas_Ppressure_PRM2017}, furthermore, has the property of being close to the onset of a semiconductor-metal transition, and hence is expected to behave like a low density electron gas. For this same reason we have also included boron doped diamond~\cite{SC:diamond:Ekimov_2004,Boeri_Diamond_PRL2004}. Lastly, we have considered high pressure sulphur hydride~\cite{DrozdovEremets_SH3_Nature2015,FloresSanna_HS3_2016EPJ,Errea_SH3_PRL2015,FloresBoeri_PerspectiveOnConvetionalHiTcSc_PhysRep2020}, which is an extreme high coupling phononic superconductor with \Tc\ of about 200~K at 200~GPa. 

The superconducting properties of the large part of these materials have been already investigated by means of first principles methods, where the electron-phonon coupling was computed from linear response density functional perturbation theory~\cite{Baroni_1987a,Baroni_DFPT_RMP2001} and Coulomb interactions were treated in RPA. In this work we have computed the Coulomb interactions with high numerical accuracy in all the considered approximation schemes (see Fig.~\ref{fig:W_materials}),
while we have taken the electron-phonon coupling values from the literature (references are listed in Tab.~\ref{tab:data}). Since we could not find ab initio electron-phonon coupling data for Nb$_3$Sn, we have carried out a full first principles study of its properties (results are collected in Appendix~\ref{app:Nb3Sn} for future reference).

\subsubsection*{Calculation of the electron-electron effective interaction} 
\begin{figure}[t] 
\begin{center}
\includegraphics[width=\columnwidth]{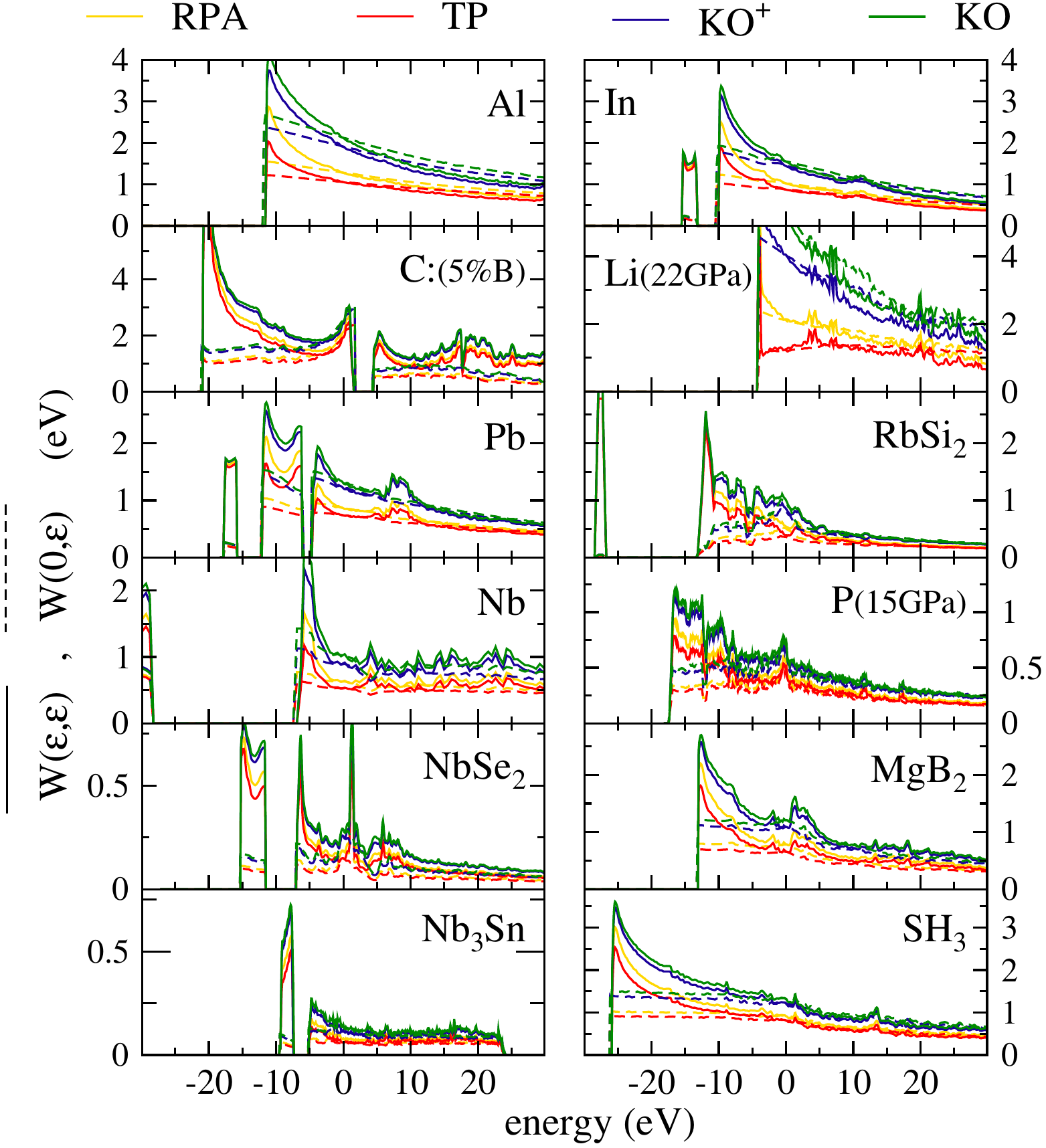} 
\caption{Coulomb interaction for the materials in our test set. Plot reports on two cuts of the $W(\varepsilon,\varepsilon')$ defined in Eq.~\ref{eq:Waverage}, a diagonal cut ($\varepsilon=\varepsilon'$) and a Fermi level cut ($\varepsilon'=0$).  }\label{fig:W_materials}
\end{center}
\end{figure}

The effective Coulomb interactions have been implemented in isotropic form (Eq.~(\ref{eq:Waverage})) in the elk FP-LAPW code~\cite{ElkCode}, which allows for the calculation of both magnetic and charge response functions. The sensitive parameters for the simulation are the \textbf{k}-point sampling of the Brillouin zone, the energy integration over empty states, which determines the accuracy of $\chi_0$, and the size of the $\chi$ matrices in \textbf{G} space. We have carried out convergence tests on all the compounds. The following results have been obtained by using at least 500 k-points per unit volume, an energy integration window up to at least 30~eV above the Fermi level and with $r_{MT}G_{max}\ge 4$ which, in LAPW codes, sets the \textbf{G} cutoff~\cite{SinghNordstrom_LAPWBook}. In Fig.~\ref{fig:W_materials} we plot two cuts of the $W(\varepsilon,\varepsilon')$ functions: a diagonal cut ($\varepsilon=\varepsilon'$) and a cut at the Fermi level $W(\varepsilon,0)$.

%%%%%%%%% RESULTS %%%%%%%%%%%%%%%
\subsection{First principle simulations of superconductors with ab initio Coulomb interactions}\label{sec:MaterialResults}
Tab.~\ref{tab:data} presents the results of our calculations for the set of superconductors in Sec.~\ref{sec:MaterialSet}. We have listed the critical temperatures obtained from Eliashberg theory by treating the Coulomb interaction at the RPA level, and the deviations ($\Delta T_c$) from these values when using the $tt$, KO$^+$ and full-KO interactions. For the computation of $T_c$ we have employed the simplified Eliashberg equations introduced in Ref.~\onlinecite{Pellegrini_SimpEliashberg_JOPM2022}. However, calculations based on density functional theory for superconductors~\cite{SPG_EliashbergSCDFT_PRL2020} provide consistent predictions.
To gain a qualitative understanding of the trend of $T_c$ across the different approximation schemes, we have computed the corresponding values of the effective Coulomb parameter $\mu$. The relative error in the RPA values of the quantity $X$ is evaluated as $\Delta X= \left(X-X^{RPA}\right)/\left[\left(X+X^{RPA}\right)/2)\right]$. 
For a comparison with the results obtained in Sec.~\ref{sec:eg}, Tab.~\ref{tab:data} also includes an estimate of the Wigner Seitz radius $r_s$ of the materials. This is defined as the $r_s$ of a HEG that has the same Fermi density of states of the material.  

\subsubsection{Analysis of the Coulomb Interaction}

\begin{figure}[t] 
\begin{center}
\includegraphics[width=\columnwidth]{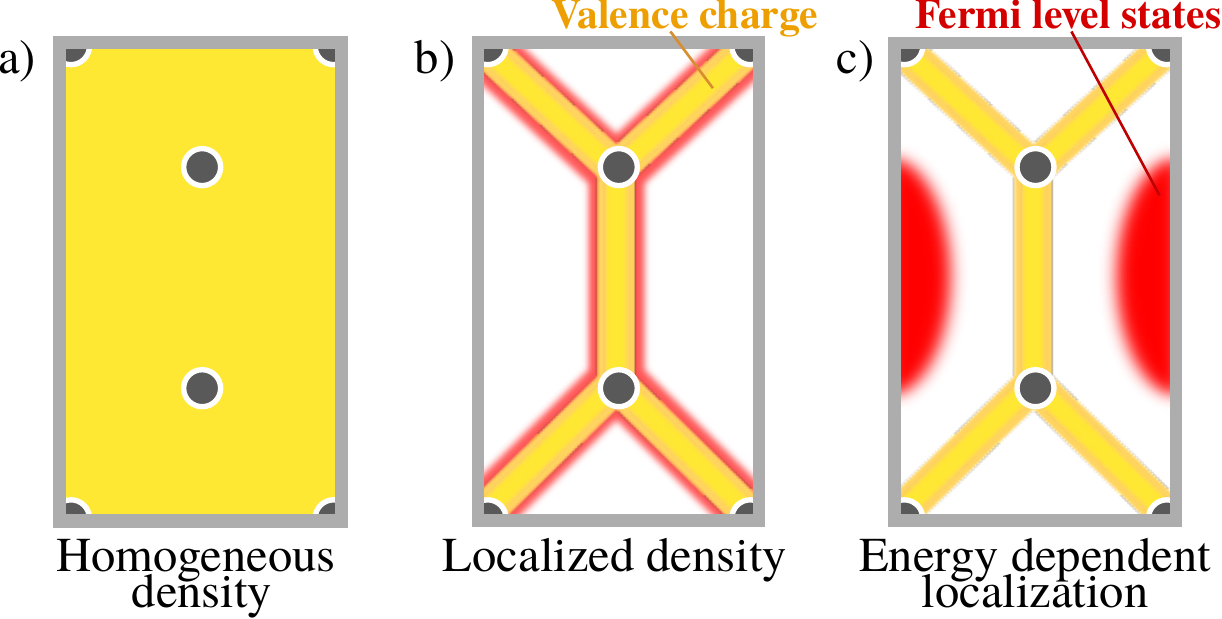} 
\caption{Three physical situations determining the role of xc corrections to Coulomb interactions:
(a) The charge density is quite homogeneous: xc effects are similar to those computed for the electron gas (see Fig.~\ref{fig:W_electronGas}). (b) The charge density is strongly localised: xc effects are reduced by the local high density environment. (c) Most of the charge density is localised but it doesn't overlap with the Fermi level orbitals: xc correlation effects are enhanced by the local low density felt by Fermi level electrons.
}\label{fig:sketch}
\end{center}
\end{figure}

We observe that the general trends of $\mu$ for the materials in Tab.~\ref{tab:data} resemble those for the HEG (upper panel of Fig.~\ref{fig:W_electronGas}), i.e., in the $tt$ scheme the Coulomb repulsion is largely reduced compared to RPA, whereas it is enhanced by assuming the KO ansatz. However, many-body corrections are in magnitude on average smaller than those expected from the electron gas for the same $r_s$, and are not strictly proportional to it. 
To explain this evidence, one must consider two aspects. First, in real materials the charge density is non-uniformly distributed since electrons are mainly localized within chemical bonds/Bloch orbitals, so that the effective screening volume may be much smaller than the cell volume $\Omega$ (see Fig.~\ref{fig:sketch}b). This is the case of strong covalent compounds like black-phosphorus and doped diamond, where the computed (large) value of $r_s$ hints at a low density behavior, but xc corrections (see the values of $\Delta \mu^{KO}$ in Tab.~\ref{tab:data}) turn out to be as small as at high density. On the other hand, $\mu$ provides a measure of the Coulomb interaction between electrons that are close to the Fermi level, and there may occur situations in which these have a poor spatial overlap with the bulk of the valence density (see Fig.~\ref{fig:sketch}c). The actual value of $r_s$ is therefore underestimated, i.e., the density felt by the electrons at the Fermi level is lower than the average density, and deviations from $\mu^{RPA}$ become sizable. 
This is the case of both lithium under pressure and Nb$_3$Sn. E.g., in lithium the states at the Fermi level have dominant p character, whereas most of the valence charge is located in s-like orbitals. 

Within our set, aluminium and SH$_3$ are certainly the two materials where Coulomb interactions more closely resemble those in a homogeneous system. This aspect can be easily seen in Fig.~\ref{fig:W_materials}, where one observes a monotonic and smooth decrease of $W$ as a function of $\varepsilon$, that is typical of the 3D electron gas~\cite{ShamKohn_screening_electrongas_PRB1966}. Consistently with this observation, we find that xc Coulomb corrections in these materials can be accurately estimated from the data of Fig.~\ref{fig:W_electronGas} at the corresponding $r_s$.

 \subsubsection{Analysis of the critical temperature}\label{sec:AnalysisTcMaterials}

Understanding how the improved description of the Coulomb interaction by the KO ansatz can impact \Tc\ is not straightforward. In fact the outcome depends on the energy structure of the Coulomb repulsion $W(\epsilon,\epsilon')$, as well as on the reduction of the latter due to retardation effects introduced by the difference between the characteristic phonon and electron energy scales~\cite{MorelAnderson_1962,ScalapinoSchriefferWilkins_StrongCouplingSC_PR1966}. According to the McMillan formula~\cite{McMillanTC}, \Tc\ is roughly determined by $\frac{\lambda}{1+\lambda}-\mu^*$, where $\mu^*$ is the reduced (Morel-Anderson) pseudopotential that accounts for the renormalization of the Coulomb interaction due to retardation effects.
We find that for aluminium the \Tc\ computed in the KO approximation is considerably lower than in RPA. This is explained by a large increase in the Coulomb repulsion ($\Delta\mu$) of about 50\% that is not mitigated by retardation effects,
which combines with a small electron-phonon coupling $\lambda$. A poor renormalization of the Coulomb repulsion due to the presence of high-frequency phonon modes is also responsible for the large \Tc\ correction in doped diamond (120 meV vibration ascribed to C-C stretching) and MgB$_2$ (60 meV E$_{2g}$ vibrational boron mode). On the other hand, in SH$_3$, that also exhibits high-frequencies hydrogen vibrations, the large $\Delta\mu$ is compensated by a strong electron-phonon coupling ($\lambda \sim 1.9$), which yields a modest reduction in \Tc\ of about 10\%. A similar result within a completely different scenario is found for indium, where $\Delta\mu$ is flattened by retardation effects associated with low-frequency phonon modes, thereby $\mu^*\ll\mu$, and $\lambda$ is rather small.
Clearly, the interplay between phonon-mediated and Coulomb interactions in determining the value of \Tc\ is complex and strongly material dependent, and the number of possible cases can not be covered by studying a limited set of materials. Nevertheless, our numerical analysis reveals a few features that are likely to be of general validity: 
The RPA treatment of the Coulomb interaction leads to a systematic overestimation of \Tc\ with respect to the KO interaction. The error in \Tc\ is in most cases of the order of 15\%, that is comparable with the uncertainty stemming from electronic structure and lattice dynamics calculations. The relative success of the RPA is largely due to error cancellation effects. In fact, improving over the dielectric screening in the $tt$ approximation leads to results that are worse than those obtained in RPA. The $tt$ approximation, despite being a formal improvement over RPA, should never be used for the calculation of superconducting properties. 
Nonetheless, the KO interaction (or at least its simplified version KO$^+$) should be preferred to the RPA in the following cases: (i) To simulate low density systems, and especially those where the electronic states close to the Fermi level are delocalized. This is the case of, e.g., electron doped semiconductors. (ii) To simulate superconductors with high phonon frequencies, especially in the weak coupling regime. In these specific cases the error associated with the use of the RPA can exceed the 40\% of \Tc.

\section{Conclusions}

State of the art ab initio methods in superconductivity have been systematically adopting the GW$^{\text{RPA}}$ approach to compute the Coulomb contribution to Cooper pairing, in spite of the fact that this approximation is not justified at metallic densities. 
In this work we have improved over the current approach by using a generalized GW self-energy, where W is given by the KO formula, that conveniently incorporates vertex corrections in the form of local field factors.

Since the KO repulsion between two electrons is stronger than the RPA, it is expected to lower the transition temperatures estimated for conventional superconductors. By using the KO ansatz with ALDA spin-symmetric and -antisymmetric xc kernels into the Eliashberg equations, we have investigated the impact of vertex corrections on the transition temperatures of twelve different metals and metallic compounds. We have found that the amount of reduction ranges from 43\% in lithium at 22 GPa pressure to 4.1\% in bulk lead, with an average reduction of 17.8\%. While these calculations employed the full KO interaction, we have introduced a simplified KO interaction containing only the spin-symmetric xc kernel $f^{+}_{xc}$, that is shown to produce nearly the same results, and can be easily implemented in existing TDDFT linear response codes. 
As a general rule, \Tc\ corrections are expected to be sizable in the weak coupling regime, for materials with high characteristic phonon frequencies and if the Fermi level charge has a small overlap with the remaining valence density, effectively leading to a low density behavior.
In these cases we recommend that the KO approximation (or at least its simplified form) replace the RPA as the optimal choice for high accuracy superconductivity simulations.

\appendix

\section{Electron phonon coupling of $\text{Nb}_{\text{3}}\text{Sn}$}\label{app:Nb3Sn}
To compile the tests in Tab.~\ref{tab:data} it is required to know the electron-phonon coupling of the material in the form of the Eliashberg spectral function ($\alpha^2F$)~\cite{AllenMitrovic1983}. These functions could be found in recent literature for all materials in the test set, apart from Nb$_3$Sn, for which we have proceeded to its calculation. In this section we report on the simulation of the $\alpha^2F$ function for  Nb$_3$Sn.

\begin{figure}[t] 
\vspace{0.2cm}
\begin{center}
\includegraphics[width=0.7\columnwidth]{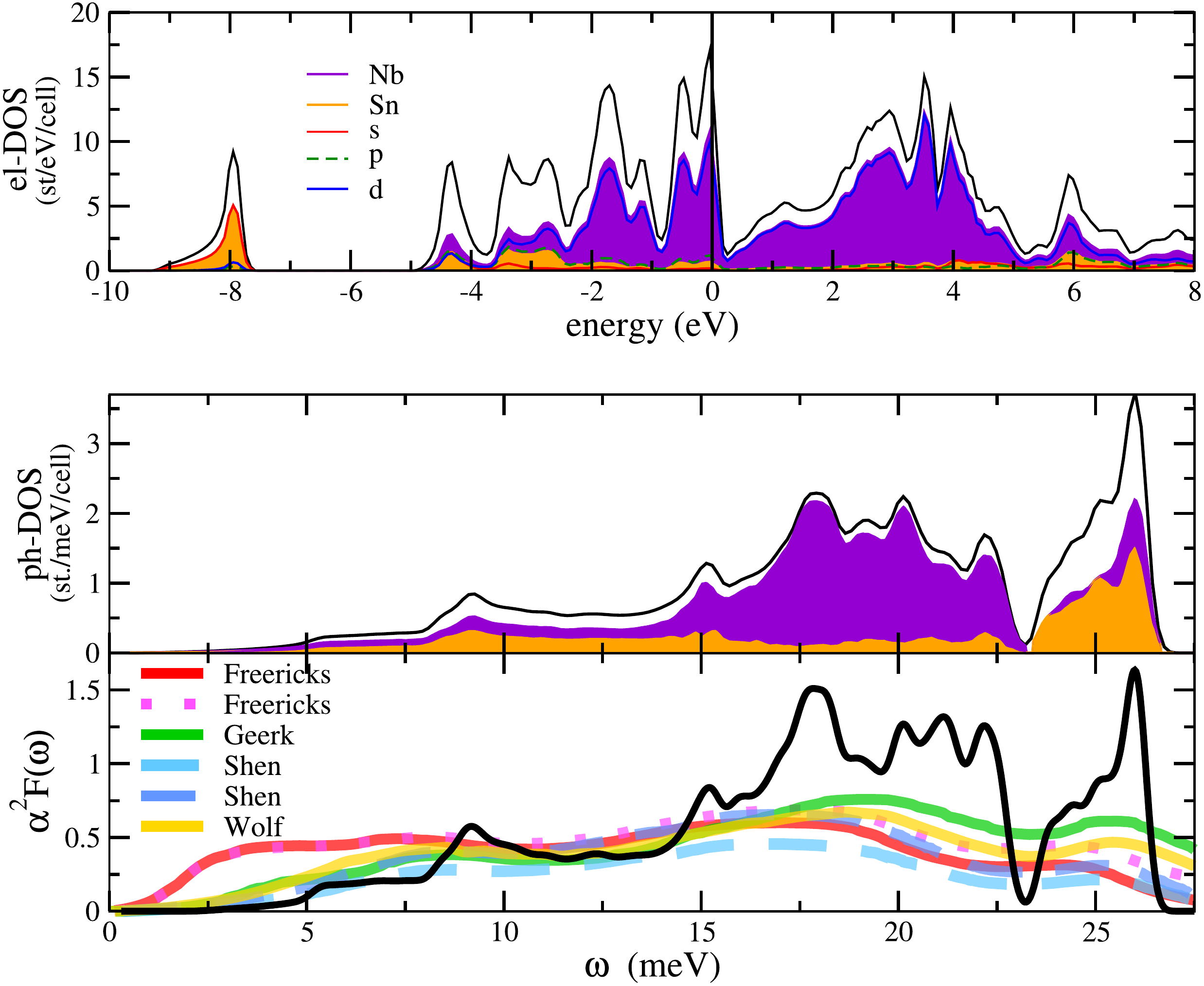} 
\caption{Electron and phonon properties of Nb$_3$Sn: Top, electronic density of states including s,p,d projection and Nb,Se projection. Center, phonon density of states with Nb, Se projection. Bottom, Eliashberg $\alpha^2F$ function (black curve) and comparison with experiments from Ref.~\onlinecite{Freericks_Nb3Sn_PRB2002,Geerk_Nb3Sn_PRB1986,Shen_Nb3Sn_PRL1972,Wolf_Nb3Sn_PRB1980}.}\label{fig:Nb3Sn}
\end{center}
\end{figure}

At room temperature Nb$_3$Sn crystallizes in the A15 crystal structure (space group $Pm\bar{3}m$, Wychoff positions 6c and 2a). Below 45~K it undergoes a cubic to tetragonal phase transition, the tetragonal distortion is very small (a/c=1.0062) and we neglect it using the A15 lattice for our simulations. We have performed all calculations with Quantum Espresso~\cite{QUANTUMESPRESSO}; the electronic structure is computed within DFT~\cite{HohenbergKohn_DFT_PR1964,KohnSham_PR1965} using the LDA approximation~\cite{PerdewZunger_LDA_1981} for the exchange correlation functional. Core states are described in the norm-conserving pseudopotential approximation and a cutoff of 70~Ry has been used for plane-wave basis set expansion. The Brillouin zone integration in the calculation of the dynamical matrices was set to a $8\times8\times8$ grid and a Methfessel-Paxton~\cite{MethfesselPaxton_PRB1989} smearing of 0.03~Ry was used. The calculated lattice parameter is $a=5.22$\AA. 
Electron phonon matrix elements are computed on a $12\times12\times12$($4\times4\times4$) \textbf{k}(\textbf{q})-grid. These are Fourier interpolated on a dense grid and then mapped on a set of 40000 \textbf{k}-points accumulated on the Fermi surface~\cite{Sanna_NbSe2_npjQM2022}, for an extremely accurate calculation of the electron phonon coupling~\cite{AllenMitrovic1983}.
The $\alpha^2F$ function, together with the electronic and phononic density of states are collected in Fig.~\ref{fig:Nb3Sn}. The  $\alpha^2F$ integrates to an electron-phonon coupling $\lambda=1.86$ and has a logarithmic averaged phonon frequency \omlog=14.5~meV. The shape of the spectral function compares well with most existing experimental estimations from tunneling inversion (also reported on the bottom panel of Fig.~\ref{fig:Nb3Sn}). However the experimental literature shows significant spread of shapes and coupling strength. The only experimental measurements in net disagreement with our simulations are the measurements from Freericks and coworkers~\cite{Freericks_Nb3Sn_PRB2002} which present extremely soft modes below 5~meV of frequency.

\bibliography{paper}
\end{document}